\documentclass[twocolumn]{aastex701}

\makeatletter
\renewcommand{\frontmatter@title@above}{%
}
\makeatother

\usepackage{amsmath}
\usepackage{amsfonts}

\begin{document}

\title{
High-energy Neutrino Predictions for T Coronae Borealis: Probing Particle Acceleration in Novae
}



\correspondingauthor{Prantik Sarmah, Sovan Chakraborty,\\ Xilu Wang}
\email{prantiksarmah@ihep.ac.cn, sovan@ittg.ac.in, wangxl@ihep.ac.cn}

\author[orcid=0000-0003-3159-7148]{Prantik Sarmah}
\affiliation{State Key Laboratory of Particle Astrophysics, Institute of High Energy Physics, Chinese Academy of Sciences, Beijing, 100049, China}
\email{prantiksarmah@ihep.ac.cn} 

\author[0000-0002-1458-8517]{Sovan Chakraborty}
\affiliation{Indian Institute of Technology Guwahati,
Guwahati, Assam-781039, India}
\email{sovan@ittg.ac.in}

\author[orcid=0000-0002-5901-9879]{Xilu Wang} 
\affiliation{State Key Laboratory of Particle Astrophysics, Institute of High Energy Physics, Chinese Academy of Sciences, Beijing, 100049, China}
\email{wangxl@ihep.ac.cn}


\begin{abstract}
The MAGIC detection of near-TeV gamma rays from the 2021 RS Oph  ($2.45$ kpc) outburst has established \textcolor{black}{recurrent novae} as TeV \textcolor{black}{particle} accelerators. \textcolor{black}{However, the origin of this emission} (hadronic vs leptonic) remains unclear due to \textcolor{black}{the lack of coincident neutrinos detected by IceCube}. The upcoming \textcolor{black}{outburst of the much closer T Coronae Borealis (T CrB, $\sim0.887$ kpc) offers a unique opportunity to detect these rare nova neutrinos.}
\textcolor{black}{Here we present the first comparative analysis of the hadronic secondary fluxes expected from the upcoming T CrB outburst and evaluate their detectability across major observatories, 
considering} two proton-acceleration mechanisms: (i) an external shock (ES)  at $\sim10^{13}$ cm, and (ii) magnetic reconnection (MR), \textcolor{black}{near the white dwarf surface at $\sim10^{9}$ cm. While the benchmark ES model predicts a gamma-ray flux detectable by current facilities, its corresponding neutrino flux largely remains undetectable}. 
\textcolor{black}{In contrast, the MR scenario generates a robust 
neutrino flux within the reach of IceCube and KM3NeT. Importantly, as the MR-produced gamma-rays are absorbed, the escaping MR neutrinos will arrive hours before any ES-origin signals. This distinct temporal separation can create a powerful phenomenological signature to disentangle the nova acceleration physics.}
\end{abstract}


\section{Introduction}
{\color{black}
Novae were long considered as GeV-scale accelerators~\citep{2014Sci...345..554A,2016ApJ...826..142C,2018A&A...609A.120F}, until the recent detection of TeV gamma-rays \citep{MAGIC:2022rmr, HESS:2022qap} from the recurrent nova RS Ophiuchi (Oph) in the Ophiuchus constellation. Typically, these novae arise in close binary systems, where the companion star's accreted hydrogen-rich material on the surface of a white dwarf undergoes a thermonuclear runaway, producing a sudden and temporary outburst. The resulting shock acceleration of protons is expected to produce secondary gamma-rays and high energy neutrinos via $pp$ interactions. Note that we refer to these gamma-rays and neutrinos as secondaries throughout this manuscript.  
RS Oph at a distance of $2.45$ kpc is a crucial example of recurrent nova, where the companion star is one massive red giant \citep{2011A&A...527A..98S,2012A&A...540A..55S}. The inter-eruption interval of recurrent novae vary over a range of $10-100$ years as per the nova system properties. 
For RS Ophiuchi (RS Oph), gamma-ray observations by MAGIC and H.E.S.S. favor a hadronic origin (inelastic $pp$ interactions) over a leptonic (inverse-Compton) scenario \citep{MAGIC:2022rmr,HESS:2022qap}, though the evidence is not yet conclusive. Hadronic interactions produce neutral and charged pions, which decay to gamma-rays and high-energy neutrinos, respectively. A coincident neutrino detection would unambiguously confirm a hadronic origin for the gamma rays. However, IceCube has not detected such correlated high-energy neutrinos from RS Oph, consistent with the low neutrino flux infereed from the observed gamma-ray emission~\citep{MAGIC:2022rmr}. Indeed, detecting such neutrinos will likely need a nova substantially closer to Earth than RS Oph.

T Coronae Borealis (T CrB) presents precisely this opportunity. Located in the constellation Coronae Borealis at a distance of only  $887^{+22}_{-23}$~pc from the Gaia Survey \citep{2021AJ....161..147B,2025ApJ...983...76H}, T CrB is a recurrent nova with past recorded eruptions in $1866$ and $1946$. This $\sim80$-year recurrence period suggests an imminent outburst in 2026~\citep{Starrfield:2025qgs,Merc:2025vkb,Petry:2025gac}. Due to its proximity, previous T CrB outbursts were nearly 10 times optically brighter ~\citep{Schaefer:2023keh} than the recent RS Oph event \citep{2010ApJS..187..275S, Zheng:2024qwt}. Thus, the upcoming T CrB eruption provides a once-in-a-lifetime opportunity to detect rare nova neutrinos \citep{IceCube:2023tuk,IceCube:2025egb,IceCube:2025gkd,Razzaque:2010kp,Bednarek:2022vey} and probe the mechanisms that produce them.

To explain nova gamma-ray emissions, multiple theoretical models have been proposed, including external, internal, and multiple shock frameworks~\citep[e.g.,][]{DeSarkar:2023nhp,Diesing:2022zkk,Steinberg:2019htp}. Among these, the external shock (ES) model—driven by the collision of the nova ejecta with the RG wind—successfully explains the multi-wavelength and gamma-ray observations of RS Oph via hadronic secondary particle production \citep{DeSarkar:2023nhp,Zheng:2022bxf}.  Alternatively, magnetic reconnection (MR) occurring within the fast nova winds near the WD surface has been proposed as another viable proton acceleration mechanism ~\citep{Bednarek:2022vey}. Crucially, these two models produce distinct multi-messenger signatures. Because MR takes place deep within the dense WD wind, the environment is expected to be completely opaque to gamma-rays created in $pp$ interactions. The weakly interacting high-energy neutrinos, however, would escape, providing a ``smoking gun" signal for this mechanism. Distinguishing between these signatures will be essential for interpreting potential high-energy emissions from T CrB.

In this work, we present the first comparative study of the high-energy secondary particle fluxes expected from the upcoming T CrB outburst under both the ES and MR particle acceleration mechanisms. 
We evaluate the detectability of these signals across major neutrino (IceCube, KM3NeT) and gamma-ray (LHAASO, Fermi-LAT, H.E.S.S., MAGIC, MACE, and HERD) observatories. As the last T CrB outburst occurred 80 years ago, precise characteristic parameters for T CrB are not available, and therefore we adopt RS Oph as our benchmark case. To account for potential variability among nova systems, we incorporate a range of observational uncertainties~\citep{Yaron:2005ys,Iijima2009,O'Brien2006,Walder2008,2003ASPC..303....9M,Cheung:2022joh}. Throughout this paper, `secondary particles' or "multi-messenger" denote the gamma rays and neutrinos resulting from hadronic $pp$ interactions. We focus primarily on TeV-scale neutrinos and the associated high-energy gamma rays above 10 GeV here.

For the conventional ES model, we find that while the gamma-ray flux should be detectable by current gamma-ray telescopes, detecting the associated neutrinos remains challenging, falling only within the upper limits of IceCube and KM3NeT. Conversely, the MR mechanism generates a neutrino flux above 1 TeV that is substantially larger than that of the ES model, making our MR benchmark nearly detectable by KM3NeT (while remaining opaque in gamma-rays). In principle, under an optimistic scenario, both ES and MR processes could operate simultaneously during the explosion, creating unique phenomenological signatures. For example, spatial separation between the ES and MR emission regions could result in a measurable time delay between their respective signals. Assuming a typical wind velocity $v_w$ of $(2000-4000)$ km/s, this delay would span several hours; for our benchmark T CrB model ($v_w = 3000$ km/s), the delay is approximately $(9-10)$ hours. Ultimately, either the multi-messenger detection or the non-detection of these correlated high-energy particles will place stringent constraints on nova acceleration mechanisms.

These rare and once in a life time phenomenological possibilities are analyzed in the following order, Sec.~\ref{sec:Ext-shock} describes the ES model, including its associated production and propagation of neutrinos and gamma rays,  as well as the detection prospects of these fluxes across various telescopes. In Sec.~\ref{sec:MR}, we detail the MR neutrino production mechanism and its detection phenomenology. Finally we conclude in Sec.~\ref{sec:conclusion}.

}

\section{External shock model}
\label{sec:Ext-shock}
 
{\color{black}
The ES model successfully explains the multiwavelength emission observed in several novae~\citep{DeSarkar:2023nhp,MAGIC:2022rmr,Zheng:2022bxf}. In the following, we briefly describe this model and its implications for proton evolution and secondary particle production.
}

\noindent
\textbf{\color{black}ES-origin gamma ray and neutrino calculations---}The ES model is assumed to be similar to the circumstellar mdedium  interaction in supernovae (SNe)~\citep{Petropoulou:2016zar}. Although the circumbinary medium in novae exhibits an asymmetric geometry, it can be effectively approximated as spherically symmetric~\citep{DeSarkar:2023nhp}.  
The number density of the circumbinary medium as a function of the shock radius, $r$, is given by, 
$n_{\rm RG} (r) = \dot{M}_{\rm RG}/{4 \pi r_{\rm i}^2 v_{\rm RG} \mu_{\rm p} m_{\rm p}}                                 *\left({r_{\rm i}}/{r} \right)^2$,
where $r_{\rm i}$ is the effective inner radius of the circumbinary medium. We take $r_{\rm i} \approx 10^{13}~\rm cm$ which is typical for  the binary separation in recurrent novae~\citep{Belczynski:1997fp}. The proton mass and the mean number of protons in the wind are denoted as $m_{\rm p}$ and $\mu_{\rm p}$, respectively. We assume $\mu_{\rm p} = 1.4$ which corresponds to $10\%$ of Helium abundance in the wind~\citep{Petropoulou:2017ymv}. For a typical RG wind with mass-loss rate, $\dot{M}_{\rm RG} = 5 \times 10^{-7}~\rm M_{\odot}/yr$ and wind velocity, $v_{\rm RG} \sim 10~\rm km/s$, $n_{\rm RG,i} \sim 10^{10} ~\rm cm^{-3}$. {\color{black} Note that, we choose  $v_{\rm RG} \sim 10~\rm km/s$ throughout our analysis of the ES model.}

Once the fast moving ejecta from the WD collide with this RG wind, the nova shock is formed and may consist of both forward and reverse fronts. 
The forward shock then propagates into the RG wind to accelerate the charged particles (electrons and protons). {\color{black} We assume the shock to be in the Sedov-Taylor (ST) phase, where  the shock radius and velocity vary with time $t$ as, $r \propto t^{2/3}$  and  $v_{\rm sh}(t)  \propto t^{-1/3}$, respectively~\citep{DeSarkar:2023nhp}. The density profile of the ejecta is assumed to vary as $r^{-3}$~\citep{MAGIC:2022rmr}.}

The energy distribution of the shock accelerated protons can be described by a power-law spectrum with an exponential cut-off. The cut-off proton energy can be obtained from the maximum energy of the accelerated protons, $E_{\rm p,max}(r)$, which depends on the acceleration timescale ($t_{\rm acc}$) and the timescales of different loss processes ($t_{\rm los}$) that hinder acceleration. The acceleration timescale in the Bohm limit is given by, $t_{\rm acc}(r) \approx 6  E_{\rm p}c/e B(r) v_{\rm sh}^2(r)$~\citep{Drury:1983}. The magnetic field, $B(r) = \sqrt{32 \pi n_{\rm RG}(r)E_{\rm th}}$, where $E_{\rm th}$ is the post-shock thermal energy with $E_{\rm th} =  \kappa_{\rm B} T_{\rm RG}$, where $\kappa_{\rm B}$ is the Boltzmann constant and $T_{\rm RG} (\sim 10^3 ~\rm K)$ is the temperature of the RG wind. For the $t_{\rm los}$, the relevant loss processes include dynamical loss and $pp$ collision loss. The dynamical loss timescale is $t_{\rm dyn}=r/v_{\rm sh}$.  The  $pp$ collision timescale is given by $t_{\rm pp} (r) = (\kappa_{\rm pp} n_{\rm RG} (r) \sigma_{\rm pp} c)^{-1}$, where $\kappa_{\rm pp} =0.5$ is the proton inelasticity and  $\sigma_{\rm pp} \approx 3 \times 10^{-26}~\rm cm^2$ is the $pp$ interaction cross-section~\citep{Kelner:2006tc}.

\begin{figure*}[htbp]
    \centering
    \begin{minipage}{\linewidth}    \includegraphics[width=0.49\linewidth]{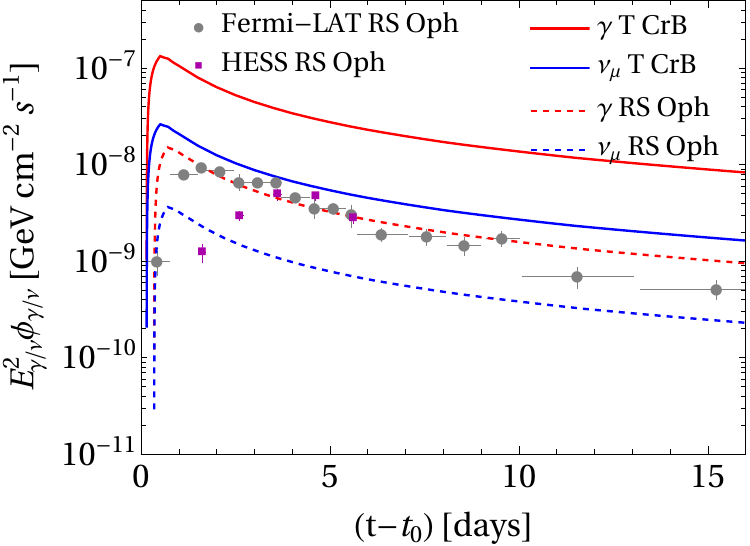}
    \includegraphics[width=0.49\linewidth]{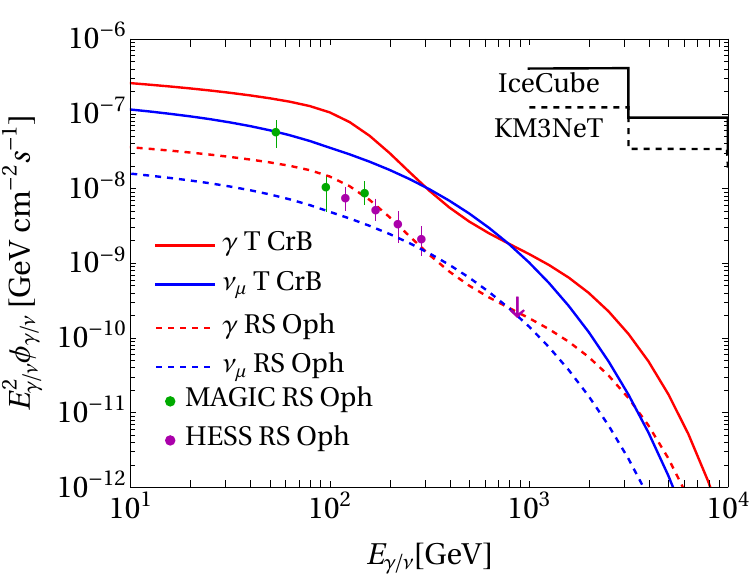}
    \end{minipage}    \caption{\textcolor{black}{\textbf{\textit{Left:}} Gamma-ray flux as a function of time (red) integrated in the energy range $50$~MeV-$500$~GeV for RS Oph using ES model. The black data points show the Fermi-LAT measurements~\cite{HESS:2022qap} in the same energy range, whereas the dark-magenta data points depict the HESS data. The Fermi-LAT data are normalised by a factor of $0.003$ as we do not consider the leptonic emission. \textbf{\textit{ Right:}} Gamma-ray (red) and muon neutrino (blue) fluxes for RS Oph at distance $2.45$~kpc. The green and dark magenta data points show the MAGIC~\citep{MAGIC:2022rmr} and HESS~\citep{HESS:2022qap} data, respectively. The black contour shows the IceCube sensitivity. The muon neutrino flux (blue curve) being far below the IceCube sensitivity explains the non-detection of neutrinos from RS Oph.  For both panel, all the model parameters are given in the benchmark values in Tab.~\ref{tab:params}, except $\epsilon_p$ which is taken to be $0.05$.}}
    \label{fig:RS_Oph_LC}
\end{figure*}

The evolution of accelerated proton number density ($N_{\rm p} (E_{_{\rm p}}, r)$) in the nova shock is governed mainly by the following equation~\citep{Petropoulou:2016zar,Sarmah:2022vra},

\begin{align}
\label{eq:protons}
\frac{\partial N_{\rm p} (E_{_{\rm p}}, r)}{\partial r} -& \frac{\partial}{\partial E_{\rm p}} \left[ \frac{E_{\rm p}N_{\rm p} (E_{_{\rm p}}, r)}{r} \right] \\ \nonumber
&+ \frac{N_{\rm p} (E_{_{\rm p}}, r)}{v_{\rm sh} (r) t_{\rm pp} (r)} = N_{\rm p}^{\rm inj} (E_{\rm p},r),    
\end{align}
with the proton injection spectrum, $N_{\rm p}^{\rm inj}(E_{\rm p},r)$ given by, $ N_{\rm p}^{\rm inj} (E_{\rm p},r) = f_{p} E_{\rm}^{-\alpha_{\rm p}} \exp{\left(-\frac{E_{\rm p}}{E_{\rm p, max} (r)} \right)}$. Here, the normalization factor $f_p$ is given by $ (9\pi/32) \epsilon_{\rm p} n_{\rm RG}(r_{\rm i}) \beta_{\rm sh}^{2} r_{\rm i}^2\times (\alpha_{\rm p}-2)$, where $\epsilon_{\rm p}$ is the fraction of the shock kinetic energy deposited in the RG wind, $\beta_{\rm sh,i}$ is the initial shock velocity, $v_{\rm sh,i}$ in unit of $c$ and $\alpha_{\rm p}$ is the power-law index. We take $\alpha_{\rm p} \geq 2$ based on the gamma-ray observation of RS Oph~\citep{MAGIC:2022rmr}.

The solution of Eq.~\eqref{eq:protons} gives the distribution of accelerated protons as a function of $E_{\rm p}$ and $r$. These accelerated protons will collide with the protons in un-shocked RG wind producing charged ($\pi^{\pm}$) and neutral ($\pi^0$) pions.   Subsequently, these pions decay to  secondaries such as electrons (positrons), neutrinos (anti-neutrinos) and gamma-rays. 
\textcolor{black}{Note that while pions can, in principle, undergo dynamical, synchrotron and hadronic interaction losses, the corresponding loss timescales are much longer than the pion decay timescale, and hence are negligible.}
Our main objective is the high energy 
neutrinos in the TeV energies, with the correlated high energy gamma-rays above $10~\rm GeV$ from these pion decays.  
Thus, we obtain the spectra of these high energy gamma-rays and neutrinos by the following differential equation~\citep{Sarmah:2022vra}, 

\begin{equation}
    \frac{\mathrm{d} N_{{\rm i}} (E_{{\rm i}},r)}{\mathrm{d}r} +\frac{N_{{\rm i}} (E_{{\rm i}},r)}{v_{\rm sh}(r) t_{\rm esc}(r)} = N_{{\rm i}}^{\rm inj} (E_{{\rm i}},r),
\end{equation}
where, $i= \nu_{\rm f}, \gamma$ and $t_{\rm esc} (r) = r/4 c$ is the escape time.  The injection spectra, $N_{\rm i}^{\rm inj}$ depend on $N_{\rm p}(E_{\rm p},r )$, $\sigma_{\rm pp}(E_{\rm p})$, $n_{\rm RG}(r)$.

\begin{table}[htbp]
\centering
\setlength{\tabcolsep}{5pt}
\begin{tabular}{lccc}
\hline\hline
\textbf{Parameter} & \textbf{Benchmark} & \textbf{Range} & \textbf{Ref.} \\
\hline
$\dot{M}_{\rm RG}~(\rm M_{\odot}\,yr^{-1})$ & $5\times10^{-7}$ & $(2$--$10)\times10^{-7}$ & [1--7] \\
$v_{\rm sh,i}~(\rm km\,s^{-1})$ & $4500$ & $3000$--$6000$ & [1--2] \\
$\alpha_{\rm p}$ & $2.2$ & $2.0$--$2.4$ & [1--2, 8--9] \\
$T_{\rm RG}~(\rm K)$ & $1500$ & $700$--$5000$ & [1--2, 9] \\
$\epsilon_{\rm p}$ & $10^{-1}$ & $(1$--$20)\times10^{-2}$ & [1, 9] \\
$L_{\rm opt}~(\rm erg\,s^{-1})$ & $10^{38}$ & $10^{37}$--$10^{39}$ & [10] \\
$T_{\rm opt}~(\rm K)$ & $8500$ & $7000$--$10000$ & [10] \\
\hline
\end{tabular}
\caption{Benchmark values and parameter ranges used in this work for ES scenario.
References [1--10] are listed in \citep{MAGIC:2022rmr,HESS:2022qap,Yaron:2005ys,Iijima2009,O'Brien2006,Walder2008,2003ASPC..303....9M,Sarmah:2022vra,DeSarkar:2023nhp,Cheung:2022joh}.}
\label{tab:params}
\end{table}

{\color{black}
Solving this evolution equation, we  obtain the gamma-ray and neutrino fluxes as,

\begin{equation}
   E_{\rm i}^{2} \phi_{\rm i}^2 (E_{\rm i}) = \frac{E_{\rm i}^2N_{\rm i}(E_{\rm i},r)}{ 4 \pi d^2 (m_{\rm e}c^2) t_{\rm esc}(r)},
   \label{eq:flux}
\end{equation}
where, $m_{\rm e}$ is the mass of electron and $d$ is the distance of the source. 
}

\noindent
{\color{black}
\textbf{Propagation effects---}In addition to the regular ES calculation, we also consider the propagation effect. The gamma-ray and neutrino fluxes given by Eq.~\ref{eq:flux} are anticipated to get attenuated during the propagation in the source ejecta and to the Earth. The dominant propagation effect for gamma-rays above 10 GeV is the pair production on the low energy photons in the source. Whereas, the weakly interacting neutrinos are not affected by such losses but may undergo flavor oscillation during the propagation. Note that we are interested in the energies above $10$~GeV only, and therefore, we do not consider low energy propagation effects, e.g., synchrotron radiation.

The attenuation of gamma-rays due to pair production can be estimated by the factor $e^{-\tau_{\rm opt}(E_{\gamma},r)}$, where the optical depth of the gamma-ray photons interacting with optical photons $\tau_{\rm opt}(E_{\gamma},r) = r \int_{\varepsilon_{\rm min}}^{\varepsilon_{\rm max}} \mathrm{d} \varepsilon n_{\rm opt}(\varepsilon, r) \sigma_{\gamma\gamma}(E_{\gamma})$~\citep{Sarmah:2022vra}. The optical photon density is $n_{\rm opt}(\varepsilon)=
f_{opt}\varepsilon^2/ (e^{\varepsilon/ \kappa_{\rm B} T_{\rm opt}}+1)$, where $T_{\rm opt}$ is the average temperature of the optical photons. The normalisation factor $f_{opt}$ is obtained from $\int \varepsilon n_{\rm opt}(\varepsilon) \mathrm{d} \varepsilon = L_{\rm opt}/(4 \pi c r^2)$, with a given optical luminosity, $ L_{\rm opt}$. Note that, absorption in the interstellar radiation field (ISRF)  during propagation to Earth is negligible below $10~\rm TeV$~\citep{Moskalenko:2005ng}.

On the other hand, the secondary neutrinos are immune to such absorption due to their weakly interacting nature, but can undergo flavour conversion due to vacuum oscillations during the propagation between the source and the earth. Note that the effect of earth matter (Mikheyev–Smirnov–Wolfenstein oscillation) is negligible for energies above $10~ GeV$, and hence neglected~\cite{Razzaque:2010kp}. The vacuum oscillation probability is effectively given by $|U|^4$, where $U$ is the Pontecorvo–Maki–Nakagawa–Sakata matrix. We use the oscillation parameters from~\citep{Esteban:2024eli} to estimate the $|U|^4$. 
}

\newblock

\newblock
\textcolor{black}{Next, we calibrate the ES model calculations plus the propagation effect described in this section, against the 2021 gamma-ray observations of RS Oph~\citep{MAGIC:2022rmr,HESS:2022qap} to obtain the benchmark values listed in Table 1, which will be adopted for the T CrB predictions in the following discussions.} The calibrated gamma-ray and neutrino fluxes for RS Oph are shown as dashed curves in Fig.~\ref{fig:RS_Oph_LC} as functions of time (left) and energy (right). In the left panel, the gamma-ray and neutrino fluxes as functions of time (up to 15 days), integrated in the energy range of 50 MeV to 500 GeV, are shown by the red and blue dashed curves, respectively. The Fermi-LAT and H.E.S.S. observational data are shown as gray and dark magenta points, respectively.
The right panel displays the gamma-ray energy spectrum (red curve) alongside flux measurements from MAGIC (green points)~\citep{MAGIC:2022rmr} and H.E.S.S. (purple points)~\citep{HESS:2022qap}. The gamma-ray flux computed using our ES model demonstrates good agreement with all observational datasets. The fluxes are plotted up to the first 16 days, which corresponds to the typical high-energy emission timescale for novae such as RS Oph~\citep{MAGIC:2022rmr,HESS:2022qap}. The small dip in the gamma-ray flux arises from absorption by the optical photons discussed above. \textcolor{black}{While \cite{MAGIC:2022rmr} reported negligible gamma-ray absorption, our adopted model parameters result in a more noticeable absorption effect. Nevertheless, as demonstrated in Fig.~\ref{fig:RS_Oph_LC}, our calculations still provide a good fit to the RS Oph gamma-ray data using these parameters.}
The right panel also plots the corresponding RS Oph neutrino flux (blue dashed) from this ES model.  Evidently, this neutrino flux is below the IceCube sensitivity (solid black)~\citep{IceCube:2018ndw} and is in agreement with the IceCube's non-observation~\citep{IceCube:2025egb,IceCube:2023tuk} of neutrino events from RS Oph.

{\color{black}
Given these calibrated parameters of the ES model, we now estimate the gamma-ray and neutrino fluxes for T CrB.    In Fig.~\ref{fig:RS_Oph_LC}, we plot these fluxes  as functions of time (left) and energy (right) by the solid red (gamma-ray) and solid blue (neutrino) curves. Clearly, these fluxes are about an order of magnitude larger than that of RS Oph due to proximity of T CrB to Earth.
}

\noindent
\textbf{\color{black}ES-origin multi-messenger signals from T CrB---}{\color{black}
The model description of the gamma-ray and neutrino fluxes in the previous section is based on our benchmark model parameters mentioned in Tab.\ref{tab:params}. There exists large uncertainties in the parameters for nova explosions listed in  Tab.\ref{tab:params}, based on multiple observations cited in the table. 
Therefore, in the following we adopt the uncertainties in parameter values in Tab.~\ref{tab:params} for T CrB (with a distance of $0.887$~kpc) to estimate the uncertainties in its signal and analyze the detection prospect of the secondary fluxes with current high energy gamma-ray and neutrino telescopes. }

\begin{figure*}[htbp]
    \centering
    \begin{minipage}{\linewidth}  
     \includegraphics[width=0.49\linewidth]{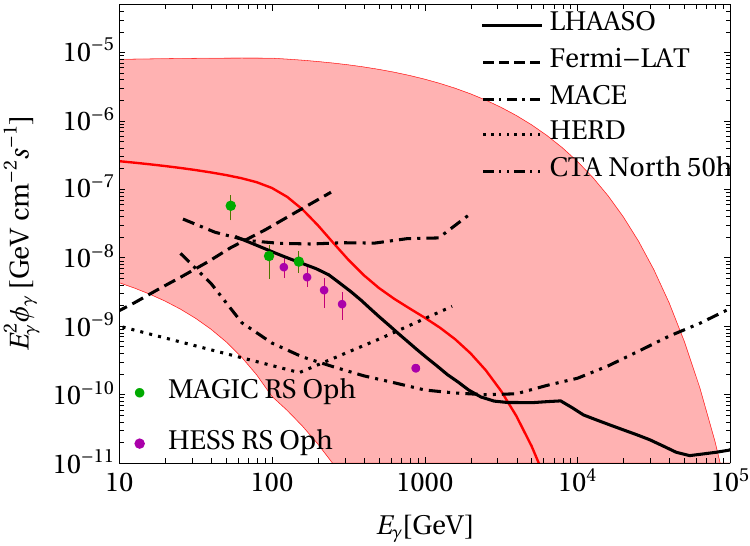}
    \includegraphics[width=0.49\linewidth]{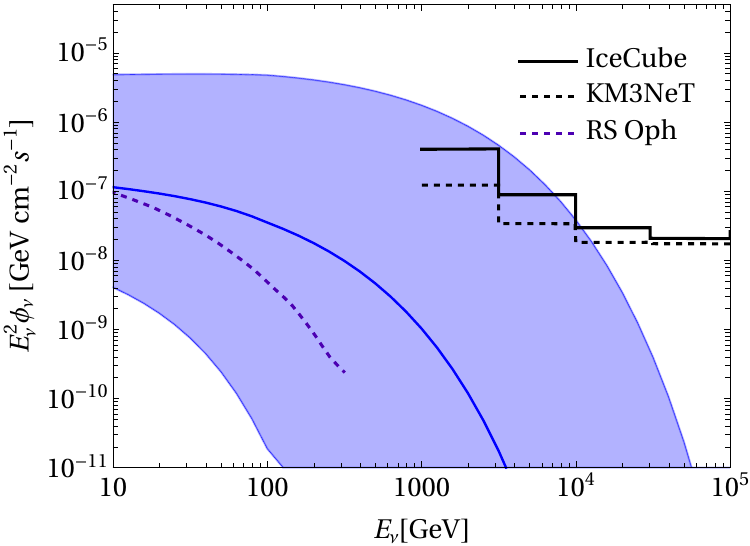}
    \end{minipage}
    \caption{\textcolor{black}{This figure shows the range of gamma-ray  and neutrino  fluxes for T CrB from the ES model.\textit{Left:} Gammma-ray flux (red curve) with model uncertainties (red band). Black lines show the sensitivities of LHAASO (solid curve), MACE (dot-dashed curve), Fermi-LAT (dashed curve) and HERD (dotted curve). The green and dark magenta data points are the observed flux of RS Ophiuchi at day 1 by MAGIC \citep{MAGIC:2022rmr} and HESS \citep{HESS:2022qap}, respectively. \textit{Right:} Muon neutrino flux (blue curve) with model uncertainties (blue bands). The purple dotted curve shows the neutrino flux estimation from RS Oph~\citep{MAGIC:2022rmr}. The solid and dotted black lines show the sensitivities of IceCube and KM3NeT, respectively. } }
    \label{fig:Flux-band}
\end{figure*}

The estimated flux of the gamma-rays and neutrinos based on the variation of model parameters are shown in the  Fig.~\ref{fig:Flux-band}. 
{\color{black} The gamma-ray (red curve) and associated uncertainties (red band) are shown in the left panel, as well as the sensitivities of current gamma-ray telescopes such as LHAASO (black solid)~\citep{Cao:2023mig}, Fermi-LAT (black dashed)~\citep{Fermi_LAT_Sensitivity}, and MACE (black dot-dashed)~\citep{Afroz:2025abo},  and two future telescopes, Cherenkov Telescope Array (CTA, black double-dot dashed) the High Energy Radiation Detection (HERD, black dotted)~\citep{HERD_Gamma_Ray_Observatory_2021,HERD:2014bpk}. }
We can see that the predicted gamma-ray flux using the benchmark model parameters (red solid curve) and indeed most of the parameter values, will be detectable by all the telescopes. In particular, with the largest sensitivity above $1$~TeV, LHAASO has the potential to probe the proton acceleration up to the highest energies, i.e., maximum proton energy $E_{\rm p,max}$ in T CrB ($E_{\rm p,max} \sim 1$~PeV). 
We also show the gamma-ray observation data of RS Oph at day $1$ by the MAGIC (green points)\citep{MAGIC:2022rmr} and HESS (dark magenta points)\citep{HESS:2022qap} telescopes in the figure. 
\textcolor{black}{
We note that the T CrB gamma-ray flux under the ES model was also recently estimated by \cite{Zheng:2024qwt}. While their calorimetric-like approach yields flux predictions consistent with ours due to overlapping parameter choices, their study focuses entirely on gamma-ray observability. Our work, by contrast, provides a broader comparative analysis of the high-energy secondary particle fluxes, in particular neutrinos, expected from the upcoming outburst, under both the ES and MR acceleration mechanisms.}

Following this validation, we show the corresponding muon neutrino flux (blue curve) and associated uncertainties (blue band) in the right panel of Fig.~\ref{fig:Flux-band}. The sensitivities of IceCube (Dec: $30^{\circ}$, \citep{IceCube:2018ndw}) and KM3NeT (Dec: $-0.01^{\circ}$, \citep{KM3NeT:2024uhg}) are plotted in black solid and black dashed, respectively.  Clearly, both the IceCube and KM3NeT will not be able to detect neutrinos for the benchmark estimate of T CrB. 
Note that we use the closest publicly available IceCube sensitivity for Dec: $30^{\circ}$, close to the declination of T CrB, Dec: $\sim 26^{\circ}$. While the sensitivity of KM3NeT is nearly constant, independent of declination \citep{KM3NeT_TDR_Part3_2015}. \textcolor{black}{Note that the sensitivities provided in these references are for $8$ and $10$ years of point source observation, respectively. For our comparison purpose, we have approximately rescaled these sensitivities for the emission time (two weeks) of the nova T CrB. Thus, this sensitivity comparison may not give the complete picture of the detection prospects.}
For a further test, we computed the number of muon neutrino events in IceCube and KM3NeT for an observation period of 15 days as:
\begin{equation}
    N_{\nu_{\mu}} = \int_{E_{\rm th}}^{\infty} \mathrm{d} E_{\nu} \phi_{\nu_{\mu}} (E_{\nu}) A_{\rm eff} (E_{\nu}),
\end{equation}
where the effective area $A_{\rm eff} (E_{\nu})$ is taken from \cite{IceCube:2016tpw} for IceCube and \cite{KM3NeT_TDR_Part3_2015} for KM3NeT. The lower limit $E_{\rm th}$ is considered to be 1 TeV for both detectors. Note that the neutrino flux from T CrB with the benchmark model is below the sensitivities of both detectors, similar to RS Oph estimate. This estimate is consistent with the non-observation of any correlated neutrino event detection in IceCube from RS Oph \citep{IceCube:2023tuk,IceCube:2025egb}. To explore the detectability of the T CrB neutrinos for both detectors,  the upper limits of the signal events $N_{\nu_{\mu}}$ that correspond to the highest neutrino flux shown in the Fig.~\ref{fig:Flux-band} are estimated as following: $N_{\nu_{\mu}} \leq 10$ events and $\leq 22$ events for IceCube and KM3NeT, respectively. Here the prediction for KM3NeT is made with the complete detector configuration, the current configuration will see fewer events. So the expected number of neutrino events is likely well below ten. Our analysis therefore indicates that the detectability of neutrinos from T CrB is highly constrained. 
Across most of the predicted flux range, the probability of detection is minimal, if not negligible. Note that, as the emission period of T CrB is expected to be a few weeks, the background neutrino events would also be significant.  Future joint gamma-ray and neutrino observations—particularly searches for gamma-ray–neutrino correlations—will provide critical constraints to reduce the large uncertainties in the current nova model parameters. Even a non-detection of neutrinos would be valuable for further constraining these models.

\section{Magnetic Reconnection Model}
\label{sec:MR}
{\color{black}
In addition to the ES model discussed above, proton acceleration via magnetic reconnection (MR) in novae has been recently proposed as a potential mechanism for high energy neutrino production \citep{Bednarek:2022vey}. Importantly, this model is expected to yield a neutrino signal entirely distinct from the ES model. In the following, we briefly describe this MR mechanism and estimate the neutrino flux for T CrB. 
}

\begin{table}[htbp]
\centering
\setlength{\tabcolsep}{10pt}
\begin{tabular}{lccc}
\hline\hline
\textbf{Parameter} & \textbf{Benchmark} & \textbf{Range} & \textbf{Ref.} \\
\hline
$L_{38}$ & $1$ & $0.5$--$5$ & [1--3] \\

$v_{\rm w,3}$ & $1$ & $0.6$--$1.2$ & [1--3] \\

$B_{7}$ & $1$ & $0.8$--$2$ & [3--4] \\

$\alpha_{\rm p}$ & $2.2$ & $2.0$--$2.4$ & [5--6] \\

$\epsilon_{\rm p,-1}$ & $1$ & $0.5$--$2$ & [3] \\

$\Omega_{-1}$ & $1$ & $0.5$--$2$ & [3] \\

$t_{10}$ & $1.4$ & $1$--$3$ & [1,5,7,8,3] \\
\hline
\end{tabular}
\caption{
Model parameters, benchmark values, and typical ranges for MR scenario.
References [1--8] correspond to
\citep{Cheung:2022joh,Aydi:2020znu,Bednarek:2022vey,2000PASP..112..873W,MAGIC:2022rmr,Guo:2014via,HESS:2022qap,Bednarek:2021hzn}.
}
\label{tab:parameters-MR}
\end{table}
\noindent
\textbf{\textcolor{black}{MR-origin neutrino calculation---}}While the magnetic fields in white dwarfs (WDs) are generally believed to be dominated by a dipole component, nova winds can drive their evolution into more complex structures~\citep{Usov,Trigilio:2004qw,Leto:2006hx,Leto:2017}. In particular, the system can be divided into two distinct regions separated by the Alfvén radius ($r_{\rm A}$). Within $r_{\rm A}$, the magnetic field retains its dipolar configuration, as the magnetic energy density exceeds the kinetic energy of the wind. While the kinetic energy of the outflow dominates beyond $r_{\rm A}$. 
Winds launched from both magnetic poles are guided by the magnetic field toward the equatorial plane, where they collide and form a hot plasma with temperatures exceeding $10^{6}$~K~\citep{Shore,Babel,Gagne:2005qd}. The resulting conditions facilitate magnetic reconnection above $r_{\rm A}$, leading to the acceleration of particles to very high energies.

The maximum energy of the accelerated protons at $r_{\rm A}$, after taking into account the energy losses due to $pp$ collisions, is given by $E_{\rm p,max} \approx 74 ~\text{TeV}~ v_{\rm w, 3}^{15/4} B_7^{1/2} / L_{38}^{3/4}$. Here, $v_{\rm w, 3}$ is the wind velocity ($v_{\rm w}$) in the unit of $3000~\rm km/s$, $B_{7}$ is the magnetic field at the WD surface in unit of $10^{7}~\rm G$, and $L_{38}$ is the luminosity of the wind in units of $10^{38}~\rm erg/s$. The Alfven radius is given by $r_{\rm A} = 2.7 R_{\rm WD} B_7^{1/2} v_{\rm w, 3}^{1/4} /L_{38}^{1/4}$, where $R_{\rm WD} = 6 \times 10^{8}~\rm cm$~\citep{1972ApJ...175..417N}.  For the distribution of the accelerated protons, we assume a power law spectrum with $E^{-\alpha_{\rm p}} \exp{\left( - E_{\rm p}/ E_{\rm p, max}\right)}$. The power law index $\alpha_{\rm p}$ is expected to be $\leq 2$ for relativistic reconnection and $>2$ for the non-relativistic case \citep{Guo:2014via,2014ApJ...783L..21S,2017ApJ...835...48B}. 
For the given magnetic field strength ($10^{7}~\rm G$) and the density of the protons ($10^{18}~\rm cm^{-3}$) close to the WD radius the magnetization parameter $\sigma = B^2/(4\pi n m c^2)$ remains small ($<<1$) resulting in non-relativistic reconnection \citep{Guo:2014via}. Hence, we choose to vary $\alpha_{\rm p }$ in the range $[2.0-2.4]$, similar to the conventional shock model discussed in Section~\ref{sec:Ext-shock}. 
Note that,   $\alpha_{\rm p }$ for non-relativistic MR is not well understood and may be significantly larger than the upper limit, $\alpha_{\rm p } = 2.4$~\citep{2010ApJ...709..963D}, resulting in heavy suppression and softer flux.

Accelerated protons that escape the reconnection region produce neutrinos and gamma-rays through the inelastic $pp$ collisions with ambient protons in the equatorial wind. Subsequent interactions of these gamma-rays with ambient thermal photons (average energy $\sim$200 eV) within the source can then generate electron-positron pairs.
The efficiency of these $\gamma-\gamma$ interactions also depends on the number density of the thermal photons, $n_{th}$. We model $n_{th}$
 with an $r^{-2}$ radial profile, normalized to $10^{19}  ~\rm cm^{-3}$  at  $R_{\rm WD}$.
As the optical depth of these thermal photons is $>> 1$, the gamma-rays would be completely absorbed in the source~\citep{Bednarek:2022vey}. While neutrinos escape the dense region without significant attenuation due to their weak interactions. The resulting neutrino flux is therefore determined by the distribution of the accelerated protons given by $N_{\rm p}(E_{\rm p}) \propto E_{\rm p}^{-\alpha_{\rm p}} \exp{\left( -E_{\rm p}/ E_{\rm p, max}\right)}$. The normalization is obtained by, $\int_{m_{\rm p}}^{\infty} E_{\rm p} N_{\rm p} (E_{\rm })\mathrm{d} E_{\rm p} = E_{\rm p, total}$. Here, $E_{\rm p, total}$ is the total energy of the accelerated protons given by $E_{\rm p, total} = 8.6 \times 10^{41} L_{38} \epsilon_{\rm p, -1} t_{10} \Omega_{-1}$ ergs, where $\epsilon_{\rm p}$ is the acceleration efficiency and $\Omega$ is the solid angle of the reconnection region, and both are taken to be of the order of $10^{-1}$. The  duration of  the reconnection process, $t$ is assumed to be similar to the duration of gamma-ray emission ($\sim 14$ days).

The interaction of these accelerated protons with the matter deposited by the WD wind produces the neutrinos. The number density of matter in this WD wind is taken as  by, $n_{\rm W} \approx 10^{18} L_{38}/(v_3^3 r^2)~\rm cm^{-3}$
. As this density is very large and the WD wind velocity being $v_{\rm W}<< c$, the accelerated protons can undergo complete cooling due to interaction with the matter.  The decay of  each charged pion ($\pi^{+}$ or $\pi^{-}$) creates two muon neutrinos, and thus, four muon neutrinos per $pp$ interaction. \textcolor{black}{However, energy losses suffered by the intermediate muons can suppress subsequent neutrino production, reducing the overall flux by a factor of two~\citep{Bednarek:2022vey}.} Therefore, for a fixed $E_{\rm p}$, the muon neutrinos yield is $N_{\nu_{\mu}} \approx 2 \mu  N_{\rm p}$, where $\mu$ is the average charged pion multiplicity per $pp$ interaction~\citep{Grosse-Oetringhaus:2009eis}.  The total number of muons neutrinos ($N_{\nu_{\mu}}^{\rm total}$) can thus be obtained by integrating the proton spectrum in the energy range, $E_{\rm p} \in [m_{\rm p}, \infty]$. 
The spectral distribution of this muon neutrino flux is obtained using the conventional  framework of Kelner et al.~\citep{Kelner:2006tc} and given by,

\begin{equation}
    \phi_{\nu_{\mu}}(E_{\nu}) \propto \int_{m_{\rm p}}^{E_{\rm p, max}} N_{\rm p}(E_{\rm p}) F_{\nu_{\mu}}(E_{\rm p},E_{\nu}) \sigma_{\rm pp}(E_{\rm p}) \mathrm{d} E_{\rm p}\,,
\end{equation}
where, $F_{\nu_{\mu}}(E_{\rm p},E_{\nu})$ is the spectral distribution function. The normalization of this muon neutrino flux can be obtained by, $\int \mathrm{d} \phi_{\nu_{\mu}}(E_{\nu}) E_{\nu} = N_{\nu_{\mu}}^{\rm total}$. The electron neutrino flux can also be obtained using the same approach and found to be smaller than the muon neutrino flux by a factor of $2$.

\begin{figure}
    \centering
    \includegraphics[width=0.95\linewidth]{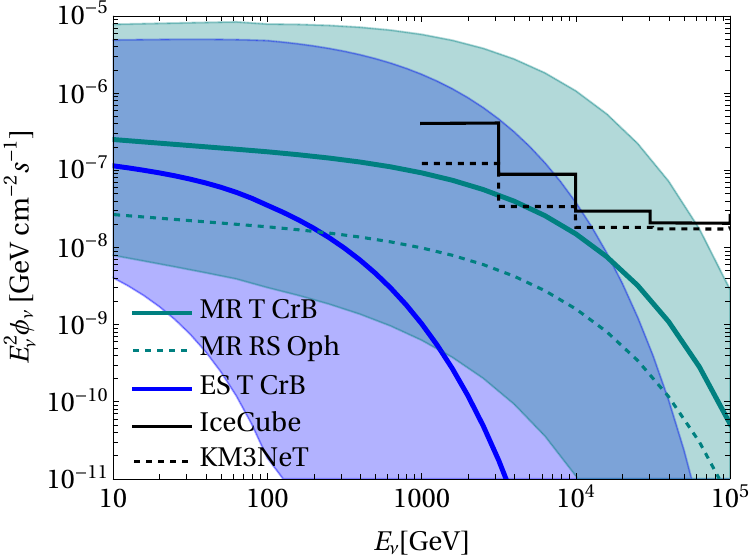}
    \caption{\textcolor{black}{Muon neutrino flux  from MR model (shown in green) compared with that from the ES model (shown in blue) for T CrB.  The  green curves depict the fluxes for T CrB (solid) and RS Oph (dotted) corresponding to the benchmark parameters in Tab.~\ref{tab:parameters-MR}, whereas the green band shows the uncertainties due to parameter variations. The sensitivities of IceCube and KM3NeT are same as in the Fig.~\ref{fig:Flux-band}.}}
    \label{fig:MR}
\end{figure}

\noindent
\textbf{{\textcolor{black}{MR-origin neutrinos predicted from T CrB---}}}The parameters responsible for the neutrino flux from magnetic reconnection are listed in Tab.~\ref{tab:parameters-MR}, with benchmark values and variation ranges provided in the second and third columns, respectively. The benchmark values are adopted primarily from Ref.~\cite{Bednarek:2022vey}, while the explored parameter ranges are motivated by observations of different novae. The references for the parameters are provided in the fourth column of Tab.~\ref{tab:parameters-MR}. Using these parameters, we compute the muon neutrino flux, taking into account the effects of neutrino oscillations as described in Section~\ref{sec:Ext-shock}. Fig.~\ref{fig:MR} shows the resulting neutrino flux for the benchmark parameter set (greenish-blue solid curve) and the range of possible fluxes (greenish-blue band).  For comparison, we also plot the muon neutrino flux from the conventional external shock model (solid blue curve with blue band) \textcolor{black}{and  the RS Oph neutrino flux  for the MR model (greenish-blue dotted curve).} The magnetic reconnection model produces a comparable flux to the external shock model at energies below $100$~GeV. At higher energies, however, the flux from magnetic reconnection is significantly larger. This flux enhancement, which can be orders of magnitude greater than that of the external shock model, leads to more optimistic detection prospects for both IceCube and KM3NeT. The corresponding sensitivity curves for IceCube (solid black line) and KM3NeT (dashed black line) are shown in Fig.~\ref{fig:MR}. Clearly, this enhanced signal would allow both detectors to probe a large portion of the parameter space for the magnetic reconnection model.  
As pointed out in Sec.~\ref{sec:Ext-shock}, we have used approximated  detector sensitivities  and they may not reveal the accurate detection scenarios. Hence, we further compute the expected number of muon neutrino events at IceCube and KM3NeT. For the benchmark case, the number of signal muon neutrinos at IceCube and KM3NeT are $\lesssim 1$ and $\lesssim 3$ events, respectively. Whereas, for the upper limit of the neutrino flux these signal events can increase up to $100$ events and can give a better prospects of detection in comparison to the ES model as the background neutrino events during the emission time remain same for both models.

{\color{black} 

Note that, for the MR mechanism to operate in novae, the WD must possess a magnetic field of $\sim 10^{6}$--$10^{8}$~G. Such strong fields are observed in isolated white dwarfs \citep[$B\sim 10^{7}$~G][]{2000PASP..112..873W}.
 In interacting binaries, 
 AM Her systems (polars) host WDs with fields of $\sim 10^{7}$--$10^{8}$~G, while intermediate polars have inferred fields of $\sim 10^{5}$--$10^{7}$~G. Recent soft X-ray modulations in Nova Herculis 2021 (V1674 Her) also imply a magnetic WD \citep{Drake:2021ncv}. However, systematic studies of WD magnetism in recurrent novae are lacking, with tentative evidence found in only a few candidates like V2487 Oph. Thus, the detection/non-detection of the MR neutrinos provides a crucial diagnostic to constrain the WD magnetic fields in recurrent novae. 
}

However, these optimistic prospects for high-energy neutrino detection in the magnetic reconnection (MR) model come at a cost: the suppression of correlated gamma-ray emission during this phase. As discussed earlier, gamma-rays cannot escape the dense MR region. Thus, any observed gamma-rays from such novae must originate from conventional external or internal shock mechanisms, either hadronic or leptonic in origin~\citep{MAGIC:2022rmr,HESS:2022qap,Cheung:2022joh,DeSarkar:2023nhp,Zheng:2022bxf}. This implies that two distinct processes—external shock (ES) and magnetic reconnection (MR)—could both contribute to the high-energy neutrino flux. Interestingly, MR neutrinos would be distinguishable from ES neutrinos by their spectral features (e.g., power-law index) and their time of onset. Because their respective emission regions are spatially disconnected, a temporal delay between these signals is expected. The characteristic length scale of the MR region is $\mathcal{O}(10^9)~\rm cm$, compared to $\mathcal{O}(10^{13})~\rm cm$ for the external shock.  Therefore, MR neutrinos are likely to arrive earlier than ES neutrinos and, crucially, will not be accompanied by correlated gamma-ray emission. 
For a typical wind velocity of $(2000-4000)$ km/sec, the delay would span several hours, depending on the specific nova system properties. In our benchmark T CrB model, assuming a wind velocity of $3000$ km/sec, the expected delay is approximately $9$ to $10$ hours. \textcolor{black}{This also suggests that MR neutrinos, being the first to reach Earth, could potentially serve as an early-warning alert for the nova explosion of T CrB}. In principle, under an optimistic scenario where both MR and ES processes operate simultaneously,  this temporal separation offers rich phenomenological possibilities for multi-messenger astronomy.

\section{Conclusion}
\label{sec:conclusion}
{\color{black}
Similar to the recent RS Oph outburst, the recurrent nova T CrB is expected to erupt soon. T CrB, being closer to the Earth, may produce a higher flux than RS Oph and thus, improving the chance of high-energy neutrino detection with detectors such as IceCube and KM3NeT. Complementing the existing studies on 1-10 GeV neutrinos and related mechanisms, this work focuses specifically on the detection prospects of the high energy neutrinos ($>1$ TeV). 
To this end, we provide the first comparative analysis of two distinct hadronic emission models for T CrB, evaluating how the detection—or strictly constrained non-detection—of these neutrinos, correlated with gamma-rays, can probe fundamental nova particle-acceleration mechanisms.

In the conventional external shock (ES) model, high-energy particles are created via collisions between accelerated protons and cold protons in the red giant wind. Calibrating our reference model to RS Oph observations and incorporating parameter variations from other novae observations yields large uncertainties (spanning roughly three orders of magnitude) in the flux predictions. Despite this, we find that the ES-derived gamma-ray flux falls well within the detection limits of Fermi-LAT, MAGIC, H.E.S.S., LHAASO,  MACE, HERD and CTA. Detecting the low-energy gamma-ray regime (e.g., with Fermi-LAT) will constrain wind density parameters and shock kinetic energy, while the highest-energy tail (e.g., with LHAASO) will probe the maximum proton energy, $E_{\rm p,max}$. In contrast, the ES model gives a pessimistic outlook for $>$ 1 TeV neutrinos; the flux drops rapidly at these energies, and only the extreme upper limits of the ES parameter space are compatible with current IceCube and KM3NeT sensitivities.

Importantly, our results highlight the magnetic reconnection (MR) model as a dominant source of high-energy neutrinos. Strong magnetic fields close to the white dwarf (WD) surface can accelerate protons to  $E_{max}\sim 10^2$~TeV—substantially higher than the ES model's limit of $\sim 1$~TeV. Thus, while the ES neutrino flux drops rapidly, the MR mechanism generates a robust TeV neutrino flux that is detectable by both IceCube and KM3NeT. Furthermore, because the dense matter and radiation environment near the WD are expected to completely absorbs MR-produced gamma-rays, any gamma-rays observed from T CrB must originate solely from the ES mechanism.

This creates a clear contrast in the expected multi-messenger signals. For the first time, we demonstrate that the spatial separation between the inner MR region and the outer ES region leads to a unique temporal signature: an arrival delay of up to several hours between the early MR-produced neutrinos and the subsequent ES-produced neutrinos and gamma-rays. This distinct time delay will provide a ``smoking gun" signature of recurrent-nova explosion mechanisms. \textcolor{black}{Furthermore, as the first emissions to reach Earth, these MR neutrinos could potentially act as an early warning for the T CrB explosion}.

Finally, we note that alternative ES models (such as internal or multiple shocks) were neglected, as their primary impact would be on the light-curve profile rather than the total time-integrated flux focused here. 
For the MR mechanism, we consider the conventional isotropic neutrino emission, as modeling this anisotropic distribution is complex and beyond the scope of this work. We also acknowledge that early-phase MR neutrinos may potentially remain undetectable under extreme models with steep spectral indices ($> 2.4$); however, even a confirmed non-detection will place stringent constraints on nova acceleration physics and the WD magnetic field.
}


\begin{acknowledgments}
The authors thank the anonymous referee for the help-ful feedback and suggestions. The work of P.S. and X.W. is supported by the National Key R$\&$D Program of China (2021YFA0718500), National Natural Science Foundation of China (Grant Nos. 1252100, 12494570, 12494574),  the Chinese Academy of Sciences (Grant No. E329A6M1) and China's Space Origins Exploration Program.
S.C. would like to acknowledge the support of DST-SERB projects CRG/2021/002961 and MTR/2021/000540.
\end{acknowledgments}





\bibliography{ref}{}
\bibliographystyle{aasjournalv7}

\end{document}